\documentclass[10pt,letterpaper]{article}
\usepackage[T1]{fontenc}
\usepackage[latin9]{inputenc}
\usepackage{geometry}
\usepackage{prettyref}
\usepackage{amsmath}
\usepackage{graphicx}
\usepackage{setspace}
\usepackage{amssymb}
\usepackage{esint}

\makeatletter
\usepackage[T1]{fontenc}
\usepackage{ae}
\usepackage{aecompl}

\usepackage[sf,md]{titlesec}

\usepackage[footnotesize,sf,hang]{caption}

\usepackage[nosort]{cite} 

\usepackage[hyperref]{collect}

\newcommand{\cpthree}{{\em CP$^\mathsf{\, 3} \! $ }}
\newcommand{\cptwo}{{\em CP$^\mathsf{\, 2}$ }}

\newcommand{\cpone}{{\em CP$^\mathsf{\, 1} \!\! $ }}

\newcommand{\rpthree}{{\em RP$^\mathsf{\, 3}$ }}
\newcommand{\rptwo}{{\em RP$^\mathsf{\, 2}$ }}

\newcommand{\adsxcp}{{\em AdS$_\mathsf{4} \times$CP$^\mathsf{\, 3} \! $ }}

\newcommand{\ssfour}{{\em S$^\mathsf{\, 2} \times \, $S$^\mathsf{\, 2}$ }}
\newcommand{\ssthree}{{\em S$^\mathsf{\, 2} \times \, $S$^\mathsf{\, 1}$ }}
\newcommand{\sseven}{{\em S$^\mathsf{\, 7} \! $ }}

\renewcommand{\vec}[1]{{\bf #1}}

\DeclareMathOperator{\tr}{tr}

\DeclareMathOperator{\re}{Re}  
\DeclareMathOperator{\im}{Im}

\DeclareMathOperator{\sech}{sech}

\AtBeginDocument{
  
}

\makeatother

\begin{document}
\begin{flushright}
BROWN-HET 1568\\
arXiv:0811.2423
\par\end{flushright}

~

\begin{doublespace}
\begin{center}
\textsf{\LARGE Giant Magnons in \adsxcp:}\\
\textsf{\LARGE Embeddings, Charges and a Hamiltonian}
\par\end{center}{\LARGE \par}
\end{doublespace}

~

\begin{onehalfspace}
\begin{center}
Michael C. Abbott and In\^{e}s Aniceto\smallskip \\
\emph{Brown University, Providence RI, USA.}\\
\emph{abbott, nes@het.brown.edu}\\
~\\
16 November 2008\\
Extended v2: ~ 23 December 2008.
\par\end{center}
\end{onehalfspace}

~

\begin{quote}
This paper studies giant magnons in $CP^{3}$, which
in all known cases are old solutions from $S^{5}$ placed into two-
and three-dimensional subspaces of $CP^{3}$, namely $CP^{1}$, $RP^{2}$
and $RP^{3}$. We clarify some points about these subspaces, and other
potentially interesting three- and four-dimensional subspaces. After
confirming that  \linebreak  $\Delta-(J_{1}-J_{4})/2$ is a Hamiltonian
for small fluctuations of the relevant `vacuum' point particle solution,
we use it to calculate the dispersion relation of each of the inequivalent
giant magnons. We comment on the embedding of finite-$J$ solutions,
and use these to compare string solutions to giant magnons in the
algebraic curve.
\end{quote}
~

\section{Introduction}

Classical string solutions in $AdS_{5}\times S^{5}$ have played an
important role in the study of the duality to $\mathcal{N}=4$ SYM.
\cite{Gubser:2002tv,Frolov:2002av,Hofman:2006xt} It seems that this
pattern is being repeated in the new $\mathcal{N}=6$ duality \cite{Aharony:2008ug},
in which planar superconformal Chern--Simons theory is dual to string
theory on $AdS_{4}\times CP^{3}$. Some of the most interesting recent
papers study strings moving in an $AdS_{2}\times S^{1}$ subspace,
where although the classical solutions are identical to those long
used in the $\mathcal{N}=4$ case, the quantum properties are different.
The results from semiclassical quantisation \cite{McLoughlin:2008ms,Alday:2008ut,Krishnan:2008zs,Arutyunov:2008if,Stefanski:2008ik,Chen:2008qq}
can be compared to those from the asymptotic Bethe ansatz, and at
present there appear to be some difficulties. \cite{Gromov:2008qe,Ahn:2008aa,Gromov:2008fy,McLoughlin:2008he,Ahn:2008tv}

This paper is instead about string solutions exploring primarily the
$CP^{3}$ factor. One would expect to find analogues of the giant
magnons \cite{Hofman:2006xt} here, which in the $\mathcal{N}=4$
case live in an $S^{2}\subset S^{5}$. And indeed, it turns out that
the same solutions exist in $CP^{3}$. \cite{Gaiotto:2008cg,Grignani:2008is}
There are two inequivalent ways to embed the basic $S^{2}$ magnon,
into either $CP^{1}=S^{2}$ or $RP^{2}=S^{2}/\mathbb{Z}_{2}$, \cite{Gaiotto:2008cg}
both two-dimensional subspaces of $CP^{3}$. 

In either theory, the anomalous dimension can be calculated as the
Hamiltonian of some spin chain. \cite{Minahan:2002ve,Minahan:2008hf,Bak:2008cp,Gaiotto:2008cg}
The giant magnons are dual to the elementary excitations of this spin
chain, and have a periodic dispersion relation $\Delta-J=\sqrt{1+f^{2}(\lambda)\sin^{2}(p/2)}$
which on the gauge side is an symptom of the discrete spatial dimension
of the spin chain, and on the string side arises from $p$ being an
angle along an equator. The conformal dimension $\Delta$ and the
R-charge $J$ are mapped by AdS/CFT to energy and angular momentum
of the string state. For the state dual to the (ferromagnetic) vacuum
of the spin chain, which is a point particle, $\Delta-J$ becomes
the Hamiltonian for small fluctuations. We confirm that in the $\mathcal{N}=6$
case, the difference $\Delta-(J_{1}-J_{4})/2$ has the same property. 

An important difference between the old $\mathcal{N}=4$ case and
the new $\mathcal{N}=6$ case is the behaviour of the function $f(\lambda)$,
the only part of the dispersion relation not fixed by supersymmetry.
\cite{Beisert:2005tm,Hofman:2006xt} In the old case, calculations
of $f(\lambda)$ at both large and small $\lambda$ give $f(\lambda)=\sqrt{\lambda}/\pi$,
and this is conjectured to be true for all $\lambda$. In the new
case, however, the function (often called $h$ instead) is $h(\lambda)=\lambda$
at small $\lambda$ but $h(\lambda)\sim\lambda^{1/2}$ at large $\lambda$.
Our knowledge of this function at large $\lambda$ comes (in both
cases) from studying classical string theory, and so depends on the
correct identification of the relevant string solutions. 

Dyonic giant magnons are those with more than one large angular momentum,
dual to a large condensate of impurities on the spin chain. These
are string solutions in $S^{3}$, and they can at least sometimes
be embedded into $CP^{3}$ in much the same way as the basic magnon,
generalising the $RP^{2}$ magnons and living in an $RP^{3}$ subspace.
\cite{Ahn:2008hj,Ryang:2008rc} There is room for dyonic solutions
with other angular momenta, truly exploring $CP^{3}$, including those
generalising the $CP^{1}$ magnon. While we have not been able to
find such solutions, we discuss where they might live. The subspace
frequently called $S^{2}\times S^{2}$ in the literature is in fact
just $RP^{2}$, and while there is a genuine $S^{2}\times S^{2}$
subspace, one cannot place arbitrary $S^{2}$ string solutions into
each factor, because the equations of motion couple the two factors.
Likewise the $S^{2}\times S^{1}$ subspace studied by \cite{Ryang:2008rc}
has extra constraints limiting what solutions can exist there.

\subsection*{Contents}

In section \ref{sec:Groups-in-ABJM} we write down a few relevant
facts about ABJM theory and its spin-chain description, and in section
\ref{sec:Geometry-of-CP3} we look at its string dual in $AdS_{4}\times CP^{3}$.
In section \ref{sec:Fluctuation-Hamiltonian} we calculate fluctuations
about the point particle solution corresponding to the spin chain
vacuum, showing that $\Delta-(J_{1}-J_{4})/2$ is a Hamiltonian for
these.

Section \ref{sec:Placing-magnons} is a catalogue of existing giant
magnon solutions in various subspaces of $CP^{3}$: single-spin magnons
in $CP^{1}$ and $RP^{2}$, and dyonic magnons in $RP^{3}$. Section
\ref{sec:Other-subspaces} looks at other subspaces of potential interest,
including the four-dimensional spaces $S^{2}\times S^{2}$ and $CP^{2}$,
and also $S^{2}\times S^{1}$. Section \ref{sec:Finite-J-corrections}
is a brief discussion of finite-$J$ solutions, which can be embedded
in the same way, and their dispersion relations.%
\footnote{Section \ref{sec:Finite-J-corrections}, and the discussion of finite
$J$ in section \ref{sec:Discussion-and-conclusion}, are new in version
2 of this paper.%
}

We discuss and conclude in section \ref{sec:Discussion-and-conclusion}.
Extra details of the geometry, and how to analyse strings in it using
Lagrange multipliers, are discussed in two appendices.

\section{Groups in ABJM theory\label{sec:Groups-in-ABJM}}

The $\mathcal{N}=6$ superconformal Chern--Simons-matter theory%
\footnote{These of theories were discovered after the explorations of 3-dimensional
superconformal theories with non-Lie-algebra guage symmetry by BLG,
\cite{Bagger:2006sk,Gustavsson:2007vu,Bagger:2007vi,VanRaamsdonk:2008ft,Gustavsson:2008bf}
and build on earlier work on Chern--Simons-matter theories by \cite{Schwarz:2004yj,Gaiotto:2007qi,Gaiotto:2008sa,Hosomichi:2008jd}.%
} of ABJM \cite{Aharony:2008ug} of interest here has gauge symmetry
$U(N)\times U(N)$. We will only study its scalars $A_{i},B_{i}$.
The fields $A_{1},A_{2}$ are matrices in the $(N,\bar{N})$ representation
of this (one fundamental index, one anti-fundamental), and the fields
$B_{1},B_{2}$ in the $(\bar{N},N)$. There is a manifest $SU(2)_{A}$
R-symmetry in which the $A$s form a doublet, and $SU(2)_{B}$ acting
on the $B$s. There is also the conformal group $SO(2,3)$, since
we are in 2+1 dimensions. Taking spacetime to be $\mathbb{R}\times S^{2}$,
we restrict attention to fields in the lowest Kaluza--Klein mode on
this $S^{2}$, i.e. in the singlet representation of $SO(3)_{r}$,
which is the spatial part of the conformal group.

In \cite{Benna:2008zy} it was proven that the full R-symmetry is
in fact $SU(4)$, with the following vector in the fundamental representation:\begin{equation}
Y^{A}=(A_{1},A_{2},B_{1}^{\dagger},B_{2}^{\dagger})\label{eq:Y-multiplet}\end{equation}
and $Y_{A}^{\dagger}$ in the anti-fundamental. If we keep only $(Y^{1},Y^{4})=(A_{1},B_{2}^{\dagger})$
then we have a subgroup called $SU(2)_{G'}$, and if we keep only
$(Y^{2},Y^{3})=(A_{2},B_{1}^{\dagger})$ then we have the subgroup
$SU(2)_{G}$.%
\footnote{These subscripts are the notation of \cite{Gaiotto:2008cg}, except
that they have $B_{1}$ and $B_{2}$ the other way around: their spin
chain vacuum is $\tr(A_{1}B_{1}^{\dagger})^{L}$ rather than the $\tr(Y^{1}Y_{4}^{\dagger})^{L}$
of \cite{Minahan:2008hf} which we use, \eqref{eq:MZ-operator-vacuum}.%
} 

This theory is dual to membranes on $AdS_{4}\times S^{7}/\mathbb{Z}_{k}$,
where $(k,-k)$ are the level numbers of the two Chern--Simons terms.
The 't Hooft limit $N\to\infty$ with $\lambda=N/k$ fixed sends $k\to\infty$,
and reduces the dual theory to type IIA strings on $AdS_{4}\times CP^{3}$. 

To find a spin-chain description, \cite{Minahan:2008hf,Gaiotto:2008cg,Bak:2008cp}
study gauge invariant operators of length $2L$ of the form\[
\mathcal{O}=\chi_{A_{1}A_{2}\cdots A_{L}}^{B_{1}B_{2}\cdots B_{L}}\tr\, Y^{A_{1}}Y_{B_{1}}^{\dagger}\; Y^{A_{2}}Y_{B_{2}}^{\dagger}\;\ldots\; Y^{A_{L}}Y_{B_{L}}^{\dagger}.\]
When $\chi$ is fully symmetric (in the $A$s, and in the $B$s) and
traceless, $\mathcal{O}$ is a chiral primary, thus protected, and
has scaling dimension $\Delta=L$. In this case the anomalous dimension,
defined $D=\Delta-L$, will be zero.

The $SU(2)\times SU(2)$ sector refers to operators $\mathcal{O}$
in which only $Y^{1}$, $Y^{2}$ and $Y_{3}^{\dagger}$, $Y_{4}^{\dagger}$
appear. (That is, only fields $A_{1}$, $A_{2}$, $B_{1}$ and $B_{2}$.
The two factors in the name are $SU(2)_{A}$ and $SU(2)_{B}$.) The
$SU(3)$ sector allows operators with $Y^{1}$, $Y^{2}$, $Y^{3}$
and $Y_{4}^{\dagger}$. For both of these, the vacuum is taken to
be \begin{equation}
\mathcal{O}_{\mathrm{vac}}=\tr\,\left(Y^{1}Y_{4}^{\dagger}\right)^{L}.\label{eq:MZ-operator-vacuum}\end{equation}
This has $\Delta=L$, and $J=L$, where $J$ is the Cartan generator
in $SU(2)_{G'}$: $J(Y^{1})=\frac{1}{2}$ and $J(Y^{4})=-\frac{1}{2}$,
thus $J(Y_{4}^{\dagger})=+\frac{1}{2}$. 

In the $SU(2)\times SU(2)$ sector, the two-loop anomalous scaling
dimension is computed by the sum of the Hamiltonians of two independent
Heisenberg $XXX$ spin chains, for the even and odd sites.  The momentum
constraint (from the $U(N)$ trace $tr$) is that the sum of their
momenta be zero. (This is slightly weaker than the $\mathcal{N}=4$
case, \cite{Minahan:2002ve} where there is one total momentum which
must be zero.)

\section{The geometry of \cpthree  \label{sec:Geometry-of-CP3} }

The string dual of ABJM theory (in the 't Hooft limit) lives in the
10-dimensional space $AdS_{4}\times CP^{3}$, with sizes specified
by the metric \begin{equation}
ds^{2}=\frac{R^{2}}{4}ds_{AdS_{4}}^{2}+R^{2}ds_{CP^{3}}^{2}\label{eq:metric-AdS-CP3-total}\end{equation}
where $R^{2}=2^{5/2}\pi\sqrt{\lambda}$. The large-$\lambda$ limit
gives strongly coupled gauge theory, dual to classical strings. In
addition to this (string-frame) metric, there is a dilaton and RR
forms, given by \cite{Aharony:2008ug}, which do not influence the
motion of classical strings.

The metric for $CP^{3}$ is given in \cite{Aharony:2008ug} as \begin{equation}
ds_{CP^{3}}^{2}=\frac{dz_{i}d\bar{z}_{i}}{\rho^{2}}-\frac{\left|z_{i}d\bar{z}_{i}\right|^{2}}{\rho^{4}},\qquad\mbox{ where }\rho^{2}=z_{i}\bar{z}_{i}\label{eq:CP3-metric-Maldacena}\end{equation}
in terms of the homogeneous co-ordinates $\vec{z}\in\mathbb{C}^{4}$,
where $\vec{z}\sim\lambda\vec{z}$ for any complex $\lambda$. The
$SU(4)$ isometry symmetry is manifest here, with $\vec{z}$ in the
fundamental representation. AdS/CFT identifies this isometry group
with the $SU(4)$ R-symmetry group, so it is natural to take $\vec{z}$
to be in the same basis as the fields $Y^{A}$ in \eqref{eq:Y-multiplet}
above. 

There are two angular parameterisations commonly used. One set of
angles was given by \cite{Pope:1984bd}: \begin{align}
ds_{CP^{3}}^{2} & =d\mu^{2}+\frac{1}{4}\sin^{2}\mu\cos^{2}\mu\left[d\chi+\sin^{2}\alpha\:\left(d\psi+\cos\theta\; d\phi\right)\right]^{2}\nonumber \\
 & \qquad+\sin^{2}\mu\left[d\alpha^{2}+\frac{1}{4}\sin^{2}\alpha\left(d\theta^{2}+\sin^{2}\theta\; d\phi^{2}+\cos^{2}\alpha\left(d\psi+\cos\theta\; d\phi\right)^{2}\right)\right]\label{eq:CP3-metric-Pope}\end{align}
with ranges $\alpha,\mu\in[0,\frac{\pi}{2}]$, $\theta\in[0,\pi]$,
$\phi\in[0,2\pi]$ and $\psi,\chi\in[0,4\pi]$. Another was given
by \cite{Cvetic:2000yp,Gibbons:1978zy}: \begin{align}
ds_{CP^{3}}^{2} & =d\xi^{2}+\frac{1}{4}\sin^{2}2\xi\left(d\eta+\frac{1}{2}\cos\vartheta_{1}\: d\varphi_{1}-\frac{1}{2}\cos\vartheta_{2}\: d\varphi_{2}\right)^{2}\nonumber \\
 & \qquad+\frac{1}{4}\cos^{2}\xi\left(d\vartheta_{1}^{2}+\sin^{2}\vartheta_{1}\: d\varphi_{1}^{2}\right)+\frac{1}{4}\sin^{2}\xi\left(d\vartheta_{2}^{2}+\sin^{2}\vartheta_{2}\: d\varphi_{2}^{2}\right)\label{eq:CP3-metric-Gibbons}\end{align}
where $\xi\in[0,\frac{\pi}{2}]$, $\vartheta_{1},\vartheta_{2}\in[0,\pi]$,
$\varphi_{1},\varphi_{2}\in[0,2\pi]$ and $\eta\in[0,4\pi]$. (This
can be obtained by building $S^{7}$ from $S^{3}\times S^{3}$ with
the seventh co-ordiante $\xi$ controlling their relative sizes.)
In appendix \ref{sec:more-metric-details} we give the maps between
these angles and the homogeneous co-ordinates.

The Penrose limit describes the geometry very near to a null geodesic
\cite{Penrose1976} and has been very important in AdS/CFT. \cite{Berenstein:2002jq}
This has been studied in $AdS_{4}\times CP^{3}$ by \cite{Gaiotto:2008cg},
where the particle travels along $\chi=4t$ with $\alpha=0$, $\mu=\pi/4$
in terms of the angles in \eqref{eq:CP3-metric-Pope}, and by \cite{Nishioka:2008gz,Grignani:2008is},
who use co-ordinates \eqref{eq:CP3-metric-Gibbons}, expanding near
$\vartheta_{1}=\vartheta_{2}=0$, $\xi=\pi/4$ with distance along
the line $\tilde{\psi}=\eta+(\varphi_{1}-\varphi_{2})/2=-2t$. In
all cases, the test particle moves along the path%
\footnote{We stress that there are not different Penrose limits for the different
giant magnon sectors. To get precisely this path $\vec{z}$, using
our conventions given in \eqref{eq:map-Gibbons-to-z} and \eqref{eq:map-Pope-to-z},
we fix in addition $\theta=\pi$ (in the first case) and $\varphi_{1}=\varphi_{2}$
(in the second), and also swop $z_{2}\leftrightarrow z_{4}$ in the
second case.%
} \begin{equation}
\vec{z}=\tfrac{1}{\sqrt{2}}\left(e^{it},\,0,\,0,\, e^{-it}\right).\label{eq:point-particle}\end{equation}
This has large angular momentum in opposite directions on the $z_{1}$
and $z_{4}$ planes, as one would expect for the state dual to the
operator \eqref{eq:MZ-operator-vacuum}. This led \cite{Minahan:2008hf}
to write this state down as the string state dual to the vacuum $\mathcal{O}_{\mathrm{vac}}$.

\section{Fluctuation Hamiltonian for the point particle\label{sec:Fluctuation-Hamiltonian}}

In the $AdS_{5}\times S^{5}$ case, the string state dual to the spin
chain vacuum $\tr(\Phi_{1}+i\Phi_{2})^{L}$ is a point particle with
$X=(\cos t,\sin t,\:0,0,\:0,0)$. This state has large angular momentum
in the 1-2 plane, $J=\Delta$. By studying small fluctuations of this
state, viewed as a string solution, one can show that $\Delta-J$
is a Hamiltonian for the physical modes. \cite{Frolov:2002av} Semiclassical
quantisation treats these modes as quantum fields with energy $\Delta-J$.
Giant magnons are exitations above this vacuum, and so their semiclassical
quantisation involves calculating quantum corrections to this energy.
\cite{Minahan:2007gf,Papathanasiou:2007gd}

In the present $AdS_{4}\times CP^{3}$ case, given the point particle
state \eqref{eq:point-particle} and the vacuum \eqref{eq:MZ-operator-vacuum},
it is reasonable to guess that $\Delta-\left(J_{1}-J_{4}\right)/2$
will play the same role. Here we confirm this, by explicitly deriving
the fluctuation Hamiltonian. 

Write the metric for the $AdS_{4}$ factor in the form\begin{align}
ds_{AdS_{4}}^{2} & =-\left(\frac{1+\vec{r}^{2}}{1-\vec{r}^{2}}\right)^{2}d\tau^{2}+\frac{4}{\left(1-\vec{r}^{2}\right)^{2}}d\vec{r}^{2}\label{eq:AdS4-metric}\end{align}
where $\vec{r}=r_{i}$, $i=1,2,3$ are zero at the centre of $AdS$,
and $\tau$ is $AdS$ time. (In our notation worldsheet space and
time are $x,t$.) For the $CP^{3}$ sector we use yet another set
of co-ordinates, which are convenient for this calculation.%
\footnote{The advantage of these co-ordinates (as opposed to the angles) is
that the identification of the charges $J_{i}$ here with those for
the magnons in section \ref{sec:Placing-magnons} and those for the
gauge theory in section \ref{sec:Groups-in-ABJM} is transparent. 

To cover the whole space with these co-ordinates we need $\beta\in[0,\pi]$
and $\epsilon\in[-1,1)$. This is clearly seen in terms of the inhomogeneous
co-ordinates $z_{1}/z_{4}=e^{i2\beta}(1+\epsilon)/(1-\epsilon)$ and
$z_{2}/z_{4}$, $z_{3}/z_{4}$. (Similar, but not identical, co-ordinates
were used by \cite{Arutyunov:2008if}.)%
} We write\begin{equation}
\vec{z}=\left(e^{i\beta}\frac{1+\epsilon}{\sqrt{2}},\: y_{1}+iy_{2},\: y_{3}+iy_{4},\: e^{-i\beta}\frac{1-\epsilon}{\sqrt{2}}\right)\label{eq:strange-new-co-ords}\end{equation}
in terms of which $\rho^{2}=\bar{z}_{i}z_{i}=1+\epsilon^{2}+\vec{y}^{2}$
(where $\vec{y}^{2}=y_{j}y_{j}$). The metric \eqref{eq:CP3-metric-Maldacena}
then becomes \begin{align*}
ds_{CP^{3}}^{2} & =\frac{(1+\epsilon^{2})d\beta^{2}+d\epsilon^{2}+d\vec{y}^{2}}{1+\epsilon^{2}+\vec{y}^{2}}\\
 & \qquad-\frac{\left(\epsilon d\epsilon+\vec{y}\cdot d\vec{y}\right)^{2}+\left(2\epsilon d\beta+y_{1}dy_{2}-y_{2}dy_{1}+y_{3}dy_{4}-y_{4}dy_{3}\right)^{2}}{\left(1+\epsilon^{2}+\vec{y}^{2}\right)^{2}}.\end{align*}

Putting these together, and dropping $R^{2}$ in \eqref{eq:metric-AdS-CP3-total}
(because we pull it out to be the action's prefactor) the full metric
becomes \begin{align}
ds^{2} & =\frac{1}{4}ds_{AdS_{4}}^{2}+ds_{CP^{3}}^{2}\label{eq:full-metric-sans-R}\\
 & =\left(-\tfrac{1}{4}-\vec{r}^{2}\right)d\tau^{2}+d\vec{r}^{2}+(1-4\epsilon^{2}-\vec{y}^{2})d\beta^{2}+d\epsilon^{2}+d\vec{y}^{2}+\ldots.\nonumber \end{align}
On the second line here we expand near $\vec{r}=\vec{y}=0$, $\epsilon=0$
and present only the terms that we will need. The point particle travels
on the line $\tau=2t$, $\beta=t$, and we define perturbations about
this as follows:\begin{align}
\tau & =2t+\tfrac{1}{\lambda^{1/4}}\tilde{\tau} & \vec{r} & =\tfrac{1}{\lambda^{1/4}}\tilde{\vec{r}}\nonumber \\
\beta & =t+\tfrac{1}{\lambda^{1/4}}\tilde{\beta} & \epsilon & =\tfrac{1}{\lambda^{1/4}}\tilde{\epsilon}\label{eq:scaled-perturbations}\\
 &  & \vec{y} & =\tfrac{1}{\lambda^{1/4}}\tilde{\vec{y}}\,.\nonumber \end{align}
The perturbations $\tilde{\tau}$ and $\tilde{\beta}$ will lead to
modes which are pure gauge, but are needed for now to maintain conformal
gauge.

The Lagrangian is $\mathcal{L}=\tfrac{1}{2}\left(-\gamma_{00}+\gamma_{11}\right)$
and the Virasoro constraints are $\gamma_{00}+\gamma_{11}=0$ and
$\gamma_{01}=0$, in terms of the induced metric $\gamma_{ab}$. The
components we need are: \begin{align*}
\gamma_{00} & =G_{\mu\nu}\partial_{t}X^{\mu}\partial_{t}X^{\nu}\\
 & =\frac{1}{\lambda^{1/4}}\left[-\partial_{t}\tilde{\tau}+2\partial_{t}\tilde{\beta}\right]\\
 & \qquad+\frac{1}{\sqrt{\lambda}}\left[-\frac{(\partial_{t}\tilde{\tau})^{2}}{4}+(\partial_{t}\tilde{\vec{r}})^{2}+(\partial_{t}\tilde{\beta})^{2}+(\partial_{t}\tilde{\epsilon})^{2}+(\partial_{t}\tilde{\vec{y}})^{2}-4\tilde{\vec{r}}^{2}-4\tilde{\epsilon}^{2}-\tilde{\vec{y}}^{2}\right]\\
 & \qquad+\frac{1}{\lambda^{3/4}}\left[-4\tilde{\vec{r}}^{2}\partial_{t}\tilde{\tau}+\partial_{t}\tilde{\beta}\left(\ldots\right)+\partial_{t}\tilde{\vec{y}}\cdot\left(\boldsymbol{\ldots}\right)\right]+o(\frac{1}{\lambda})\end{align*}
where $\left(\ldots\right)$ indicates terms not needed for this calculation,
and\begin{align*}
\gamma_{11} & =G_{\mu\nu}\partial_{x}X^{\mu}\partial_{x}X^{\nu}\\
 & =\frac{1}{\sqrt{\lambda}}\left[-\frac{(\partial_{x}\tilde{\tau})^{2}}{4}+(\partial_{x}\tilde{\vec{r}})^{2}+(\partial_{x}\tilde{\beta})^{2}+(\partial_{x}\tilde{\epsilon})^{2}+(\partial_{x}\tilde{\vec{y}})^{2}\right]+o(\frac{1}{\lambda}).\end{align*}

Next we define the string's conserved charges. $\Delta$ is the charge
generated by time translation: \begin{align*}
\Delta & =2\sqrt{2\lambda}\int dx\:\frac{\partial\mathcal{L}\left[\tau,\vec{r},\beta,\epsilon,\vec{y}\right]}{\partial\,\partial_{t}\tau}\\
 & =2\sqrt{2}\,\lambda^{3/4}\int dx\:\frac{\partial\tilde{\mathcal{L}}\left[\tilde{\tau},\tilde{\vec{r}},\tilde{\beta},\tilde{\epsilon},\tilde{\vec{y}}\right]}{\partial\,\partial_{t}\tilde{\tau}}\end{align*}
and $J_{i}$ is the charge generated by rotation of the $z_{i}$ complex
plane:%
\footnote{Note that in deriving these charges we treat $Z_{1},...,Z_{4}$ as
independent fields, even though they are in fact related through $\vec{Z}\sim\lambda\vec{Z}$,
which defines $CP^{3}$ from $\mathbb{C}^{4}$. Therefore, we do this
before adopting the parametrisation \eqref{eq:strange-new-co-ords},
in which we have fixed some of this gauge freedom by writing only
six (not eight) real co-ordinates.%
} \begin{align}
J_{1} & =2\sqrt{2\lambda}\int dx\:\frac{\partial\mathcal{L}}{\partial\,\partial_{t}(\arg Z_{1})}\nonumber \\
 & =2\sqrt{2\lambda}\int dx\left[\frac{\im\left(\bar{Z}_{1}\partial_{t}Z_{1}\right)}{\rho^{2}}-\frac{\left|Z_{1}\right|^{2}\sum_{i}\im\left(\bar{Z}_{i}\partial_{t}Z_{i}\right)}{\rho^{4}}\right]\label{eq:Defn-Jz1-with-rho}\end{align}
\vspace{-1em}\begin{align*}
J_{4} & =2\sqrt{2\lambda}\int dx\left[\frac{\im\left(\bar{Z}_{4}\partial_{t}Z_{4}\right)}{\rho^{2}}-\frac{\left|Z_{4}\right|^{2}\sum_{i}\im\left(\bar{Z}_{i}\partial_{t}Z_{i}\right)}{\rho^{4}}\right].\end{align*}
Substituting in the above mode definitions, we get\begin{align}
\Delta & =\sqrt{2}\int dx\left[\sqrt{\lambda}+\frac{\lambda^{1/4}}{2}\partial_{t}\tilde{\tau}+4\tilde{\vec{r}}^{2}+o(\frac{1}{\lambda^{1/4}})\right]\label{eq:Delta-from-lagrangean}\\
J_{1} & =\sqrt{2}\int dx\left[\sqrt{\lambda}+\lambda^{1/4}\partial_{t}\tilde{\beta}-4\tilde{\epsilon}^{2}-\tilde{\vec{y}}^{2}+\left(\tilde{y}_{2}\partial_{t}\tilde{y}_{1}-\tilde{y}_{1}\partial_{t}\tilde{y}_{2}+\tilde{y}_{4}\partial_{t}\tilde{y}_{3}-\tilde{y}_{3}\partial_{t}\tilde{y}_{4}\right)+o(\frac{1}{\lambda^{1/4}})\right]\nonumber \\
J_{4} & =\sqrt{2}\int dx\left[-\sqrt{\lambda}-\lambda^{1/4}\partial_{t}\tilde{\beta}+4\tilde{\epsilon}^{2}+\tilde{\vec{y}}^{2}+\left(\tilde{y}_{2}\partial_{t}\tilde{y}_{1}-\tilde{y}_{1}\partial_{t}\tilde{y}_{2}+\tilde{y}_{4}\partial_{t}\tilde{y}_{3}-\tilde{y}_{3}\partial_{t}\tilde{y}_{4}\right)+o(\frac{1}{\lambda^{1/4}})\right].\nonumber \end{align}
These diverge as $\lambda\to\infty$, but for the linear combination
used below, the $o(\sqrt{\lambda})$ terms cancel. The $o(\lambda^{1/4})$
terms, linear in the fluctuations, can be re-written as quadratic
$o(1)$ terms using the Virasoro constraint $\gamma_{00}+\gamma_{11}=0$.
This leads to\begin{align*}
 & \Delta-\frac{J_{1}-J_{4}}{2}\\
 & \qquad=\frac{\sqrt{2}}{2}\int dx\Bigg[(\partial_{t}\tilde{\vec{r}})^{2}+(\partial_{x}\tilde{\vec{r}})^{2}+4\tilde{\vec{r}}^{2}+(\partial_{t}\tilde{\epsilon})^{2}+(\partial_{x}\tilde{\epsilon})^{2}+4\tilde{\epsilon}^{2}+(\partial_{t}\tilde{\vec{y}})^{2}+(\partial_{x}\tilde{\vec{y}})^{2}+\tilde{\vec{y}}^{2}\\
 & \qquad\qquad\qquad\qquad-\frac{(\partial_{t}\tilde{\tau})^{2}}{4}-\frac{(\partial_{x}\tilde{\tau})^{2}}{4}+(\partial_{t}\tilde{\beta})^{2}+(\partial_{x}\tilde{\beta})^{2}\Bigg]+o(\frac{1}{\lambda^{1/4}}).\end{align*}
The terms on the last line are the gauge modes, generating infinitesimal
reparameterisations, so would not be included in semiclassical quantisation.
After dropping these, we are left with the Hamiltonian%
\footnote{This $\mathcal{H}$ is the two-dimensional Hamiltonian that one would
obtain from the quadratic part of the fluctuation Lagrangian $\mathcal{L}=\tfrac{1}{2}(-\gamma_{00}+\gamma_{11})$
by naively dropping terms linear in time derivative and reversing
the signs of the terms quadratic in the time derivative. But note
that without dropping these $o(\lambda^{1/4})$ terms, the string
Hamiltonian is fixed to zero by the Virasoro constraint $\gamma_{00}+\gamma_{11}=0$,
which we have used to derive $\mathcal{H}$.%
} $\Delta-\frac{J_{1}-J_{4}}{2}=\sqrt{2}\int dx\,\mathcal{H}$, where%
\footnote{The obvious charges one could add to $\Delta-\left(J_{1}-J_{4}\right)/2$,
while keeping it finite, are $J_{2}$ and $J_{3}$. These will add
terms like $\tilde{y}_{2}\partial_{t}\tilde{y}_{1}-\tilde{y}_{1}\partial_{t}\tilde{y}_{2}$
to $\mathcal{H}$.%
}\[
\mathcal{H}=\frac{1}{2}\left[\:(\partial_{t}\tilde{\vec{r}})^{2}+(\partial_{x}\tilde{\vec{r}})^{2}+4\tilde{\vec{r}}^{2}\;+\;(\partial_{t}\tilde{\epsilon})^{2}+(\partial_{x}\tilde{\epsilon})^{2}+4\tilde{\epsilon}^{2}\;+\;(\partial_{t}\tilde{\vec{y}})^{2}+(\partial_{x}\tilde{\vec{y}})^{2}+\tilde{\vec{y}}^{2}\:\right].\]
This describes eight massive modes: the three $\tilde{r}_{i}$ in
$AdS_{4}$, plus $\tilde{\epsilon}$ and the four $\tilde{y}_{i}$
in $CP^{3}$. As was noted by \cite{Arutyunov:2008if}, one of the
$CP^{3}$ modes, $\tilde{\epsilon}$, has reached across the aisle
to have the same mass as the $AdS$ modes $\tilde{\vec{r}}$. The
same list of masses was also found by \cite{Nishioka:2008gz,Gaiotto:2008cg,Grignani:2008is}
when studying the Penrose limit, and by \cite{McLoughlin:2008ms,Alday:2008ut,Krishnan:2008zs,Arutyunov:2008if}
for modes of spinning strings in the $AdS_{2}\times S^{1}$ subspace.

\section{Placing giant magnons into \cpthree\emph{ \label{sec:Placing-magnons}}}

Recall that the Hoffman--Maldacena giant magnon \cite{Hofman:2006xt}
is a rigidly rotating classical string solution in $\mathbb{R}\times S^{2}$,
given in timelike conformal gauge by\begin{align}
\cos\theta_{\mathrm{mag}} & =\sin\frac{p}{2}\sech u\label{eq:basic-magnon}\\
\tan\left(\phi_{\mathrm{mag}}-t\right) & =\tan\frac{p}{2}\tanh u\nonumber \end{align}
where $u=(x-t\:\cos\tfrac{p}{2})/\sin\tfrac{p}{2}$ is the boosted
spatial co-ordinate for a soliton with worldsheet velocity $\cos(p/2)$.
The spacetime is $ds^{2}=-dt^{2}+d\theta^{2}+\sin^{2}\theta\: d\phi^{2}$
--- by timelike gauge we mean that the target-space time is also worldsheet
time.%
\footnote{What we call timelike conformal gauge is sometimes called static conformal
gauge. In our conventions, $AdS$ time $\tau$ is given by $\tau=2t$.
However, because of the factor  $\frac{1}{4}$ in the metric \eqref{eq:full-metric-sans-R},
it is $t$ rather than $\tau$ which is physical time.%
} 

We define conserved charges here as follows:\begin{align}
\Delta & =\sqrt{2\lambda}\int dx\:1\label{eq:defn-Delta-timelike-gauge}\\
J_{\mathrm{sphere}} & =\sqrt{2\lambda}\int dx\:\im\left(\bar{W}_{1}\partial_{t}W_{1}\right).\label{eq:defn-J-sphere}\end{align}
This $\Delta$ matches \eqref{eq:Delta-from-lagrangean} used above
when the $AdS$ fluctuations $\tilde{\tau}$ and $\tilde{\vec{r}}$
are turned off. Note that we keep the same prefactor $\sqrt{2\lambda}$
here, which is not the one we would use in the $AdS_{5}\times S^{5}$
case. Finally, we write the complex embedding co-ordinates $W_{1}=e^{i\phi_{\mathrm{mag}}}\sin\theta_{\mathrm{mag}}$
and $W_{2}=\cos\theta_{\mathrm{mag}}$.%
\footnote{Our notation is that $(w_{1},w_{2})$ are complex embedding co-ordinates
for the sphere, while $z_{i}$ are for $CP^{3}$. Capital letters
indicate a string solution in this space.%
}

Both $\Delta$ and $J_{\mathrm{sphere}}$ are infinite for the solution
\eqref{eq:basic-magnon}, but their difference is finite:\begin{align*}
\Delta-J_{\mathrm{sphere}} & =2\sqrt{2\lambda}\:\sin\left(\frac{p}{2}\right).\end{align*}
The parameter $p$ is the (absolute value of the) momentum of the
spin chain excitation in the dual gauge theory, which is why this
is called a dispersion relation. It is also equal to the opening angle
$\Delta\phi_{\mathrm{mag}}$ of the string solution on the equator
$\theta_{\mathrm{mag}}=\tfrac{\pi}{2}$.

We now turn to solutions in $\mathbb{R}\times CP^{3}$, with metric
$ds^{2}=-dt^{2}+ds_{CP^{3}}^{2}$. All solutions will be in conformal
gauge, and with worldsheet time $t$ related to $AdS$ time $\tau$
by $\tau=2t$, so we will continue to use the definition of $\Delta$
from \prettyref{eq:defn-Delta-timelike-gauge}, although for $J$
we must now use \eqref{eq:Defn-Jz1-with-rho}. We will also continue
to use the parameter $p\in[0,2\pi]$ in all the cases below, and while
this should still be a momentum in the dual theory, we make no comment
here on the precise factors involved.

\subsection{The subspace \cpone  \label{sub:Subspace-CP1}}

If we set $z_{2}=z_{3}=0$, or in terms of angles \eqref{eq:CP3-metric-Pope},
$\alpha=0$, then we obtain the space $CP^{1}=S^{2}$ with metric\begin{equation}
ds^{2}=\frac{1}{4}\left[d(2\mu)^{2}+\sin^{2}(2\mu)d\!\left(\frac{\chi}{2}\right)^{2}\right].\label{eq:S2-alpha-zero}\end{equation}
This is a sphere of radius $\frac{1}{2}$, so to place the magnon
solution \eqref{eq:basic-magnon} here (as was done by \cite{Gaiotto:2008cg})
maintaining conformal gauge we need to set\begin{align}
2\mu & =\theta_{\mathrm{mag}}(2x,2t)\label{eq:2x-2t-scaling}\\
\frac{\chi}{2} & =\phi_{\mathrm{mag}}(2x,2t)\,.\nonumber \end{align}
Using the map \eqref{eq:map-Pope-to-z}, given in appendix \ref{sec:more-metric-details},
and choosing $\theta=\pi$, we obtain \begin{align}
\vec{Z}(x,t) & =\frac{1}{\sqrt{2}}\left(e^{\frac{i}{2}\phi_{\mathrm{mag}}(2x,2t)}\sqrt{1-\cos\theta_{\mathrm{mag}}(2x,2t)}\,,\:0,\:0,\: e^{-\frac{i}{2}\phi_{\mathrm{mag}}}\sqrt{1+\cos\theta_{\mathrm{mag}}}\right)\label{eq:GGY-magnon-Z}\\
 & =\left(e^{it+f(2u)}\sin\frac{\theta_{\mathrm{mag}}(2x,2t)}{2}\,,\:0,\:0,\: e^{-it-f(2u)}\cos\frac{\theta_{\mathrm{mag}}(2x,2t)}{2}\right).\nonumber \end{align}

Calculating charges for this solution, using definitions \eqref{eq:Defn-Jz1-with-rho}
for $J$ and \eqref{eq:defn-Delta-timelike-gauge} for $\Delta$,
we recover the dispersion relation%
\footnote{Note that if you were to omit the second term in \eqref{eq:Defn-Jz1-with-rho}
when calculating $J$, thus effectivly using \eqref{eq:defn-J-sphere}
appropriate for the sphere, you would get instead $\Delta-(J_{1}-J_{4})/2=\sqrt{2\lambda}\, p\cos\left(\frac{p}{2}\right)$.
In the $RP^{2}$ and $RP^{3}$ subspaces discussed below, this second
term vanishes. \label{fn:Note-on-wrong-J-defn}%
} \begin{equation}
\Delta-\frac{J_{1}-J_{4}}{2}=\sqrt{2\lambda}\:\sin\left(\frac{p}{2}\right).\label{eq:GGY-disp-rel}\end{equation}

We should check that this subspace is a legal one, meaning that solutions
found here are guaranteed to be solutions in the full space. This
can be done by finding the conformal gauge equations of motion coming
from the Polyakov action with the metric \eqref{eq:CP3-metric-Pope},
and confirming that $\alpha$'s equation is solved by $\alpha=0$.%
\footnote{In addition to solving the conformal gauge equations of motion, a
string solution must be in conformal gauge, i.e. must solve the Virasoro
constraints. If the solution on the subspace is in conformal gauge,
then it follows trivially that the solution in the full space is too:
the induced metric $\gamma_{ab}=\partial_{a}X^{\mu}\partial_{b}X^{\nu}G_{\mu\nu}$
is influenced only by those directions the solution explores, and
in these directions the metric $G_{\mu\nu}$ is the same in both the
full space and the subspace.%
} But in this case it is easier to note that $z_{2}=z_{3}=0$ trivially
solves their equations of motion, \eqref{eq:eq-of-m-in-Z}, which
we derive in appendix \ref{sec:strings-sans-trig}.

\subsection{The subspace \rptwo\label{sub:Subspace-RP2}}

A second embedding of the $S^{2}$ solution was first used by \cite{Grignani:2008is}%
\footnote{We discuss the equations of motion used by \cite{Grignani:2008is}
for strings in $CP^{3}$ in appendix \ref{sub:Constraining-S7-solutions}. %
} \begin{equation}
\vec{Z}(x,t)=\frac{1}{\sqrt{2}}\left(e^{i\phi_{\mathrm{mag}}(x,t)}\sin\theta_{\mathrm{mag}}(x,t)\,,\:\cos\theta_{\mathrm{mag}}\,,\:\cos\theta_{\mathrm{mag}}\,,\: e^{-i\phi_{\mathrm{mag}}}\sin\theta_{\mathrm{mag}}\right).\label{eq:GHO-magnon-Z}\end{equation}

\begin{figure}
\begin{centering}
\includegraphics[height=5.5cm]{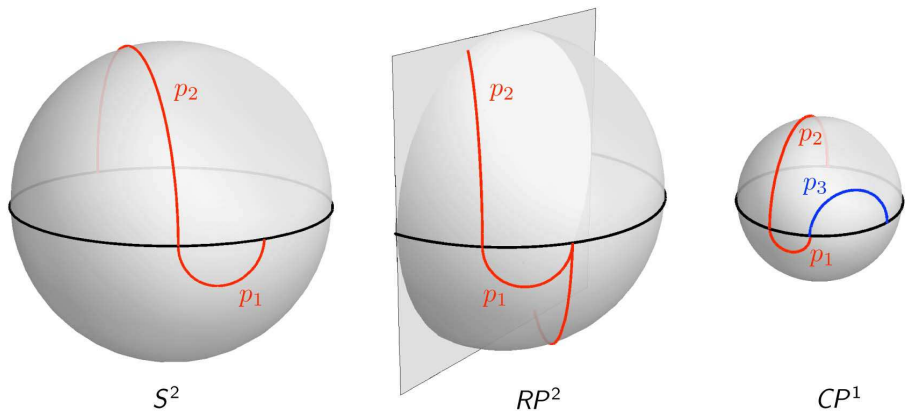}
\par\end{centering}

\caption{Two giant magnons are shown (in red) on the unit sphere $S^{2}$ (left),
on $RP^{2}$ (centre, drawn here as half a sphere) and on $CP^{1}$,
a sphere of radius $\frac{1}{2}$ (right). In all cases they have
$p_{1}=\tfrac{1}{2}$ and $p_{2}=\pi-\tfrac{1}{2}$, which leads to
a closed string in the $RP^{2}$ case, but not in the $S^{2}$ or
$CP^{1}$ cases. \label{fig:Two-magnons} \smallskip\smallskip \protect \\
In both the $RP^{2}$ and $CP^{1}$ cases, the equator is of length
$\pi$, and we parameterise it by $\beta\in[0,\pi]$. The magnon with
$p_{1}=\frac{1}{2}$ spans $\Delta\beta=\frac{1}{2}$ in the $RP^{2}$
case, but only $\Delta\beta=\frac{1}{4}$ in the $CP^{1}$ case. On
$CP^{1}$ we have also drawn a third magnon (in blue) with $p_{3}=1$,
which spans the same length of equator $\Delta\beta=\frac{1}{2}$
as does the $p_{1}$ magnon on $RP^{2}$.}

\end{figure}

This solution lives in an $RP^{2}$ subspace, as can be seen by simply
rotating some of the planes in $\mathbb{C}^{4}=\mathbb{R}^{8}$ by
$\frac{\pi}{4}$: in terms of new co-ordinates $\vec{w}$ defined
by \begin{align}
w_{1} & =\tfrac{1}{\sqrt{2}}\left(z_{1}+\bar{z}_{4}\right) & w_{4} & =\tfrac{1}{\sqrt{2}}\left(z_{1}-\bar{z}_{4}\right)\label{eq:rotate45-defn-w}\\
w_{2} & =\tfrac{1}{\sqrt{2}}\left(z_{2}+\bar{z}_{3}\right) & w_{3} & =\tfrac{1}{\sqrt{2}}\left(z_{2}-\bar{z}_{3}\right),\nonumber \end{align}
this solution has $w_{3}=w_{4}=0$ and is precisely the original giant
magnon in the other two co-ordinates: \[
(W_{1},W_{2})=\left(e^{i\phi_{\mathrm{mag}}}\sin\theta_{\mathrm{mag}}\,,\:\cos\theta_{\mathrm{mag}}\right).\]
The reason this is $RP^{2}$ rather than $S^{2}$ is that sending
$(w_{1},w_{2})\to-(w_{1},w_{2})$ gives an overall sign change on
$\vec{z}$, and these two points are identified in $CP^{3}$.%
\footnote{In $S^{2}$, the standard co-ordinates have ranges $\theta\in[0,\pi]$
and $\phi\in[0,2\pi]$, and changing $\theta\to\pi-\theta$ and $\phi\to\phi+\pi$
simultaneously moves you to the antipodal point on $S^{2}$. But performing
this change in the subspace of $CP^{3}$ parameterised by \eqref{eq:GHO-magnon-Z}
changes $\vec{z}\to-\vec{z}$, and these two points are identified
by the definition of $CP^{3}$. This is what makes the subspace $RP^{2}=S^{2}/\mathbb{Z}_{2}$
instead of $S^{2}$. To obtain co-ordinates which cover this subspace
only once, we can shorten the range of either $\theta$ or $\phi$,
and in figure \ref{fig:Two-magnons} we choose to restrict to $\phi\in[0,\pi]$
while keeping $\theta\in[0,\pi]$.%
}

The subspace which this magnon explores can also be obtained from
the metric \eqref{eq:CP3-metric-Gibbons}, by fixing $\vartheta_{1}=\tfrac{\pi}{2}$,
$\vartheta_{2}=\tfrac{\pi}{2}$, $\varphi_{1}=0$ and $\eta=0$. The
metric then becomes \[
ds^{2}=d\xi^{2}+\sin^{2}\xi\: d\!\left(\frac{\varphi_{2}}{2}\right)^{2}\]
and the magnon \eqref{eq:GHO-magnon-Z} is simply $\xi=\theta_{\mathrm{mag}}(x,t)$,
$\varphi_{2}=2\phi_{\mathrm{mag}}(x,t)$. This can be checked to be
a legal restriction from the equations of motion for the four angles
fixed. 

This subspace is sometimes, rather misleadingly, referred to as $S^{2}\times S^{2}$.
It is true that $\left|z_{1}\right|^{2}+\left|z_{2}\right|^{2}=\tfrac{1}{2}$
and $\left|z_{3}\right|^{2}+\left|z_{4}\right|^{2}=\tfrac{1}{2}$,
and $\im z_{2}=0=\im z_{3}$. These restrictions alone would describe
a subspace of $\mathbb{C}^{4}$, namely $S^{2}\times S^{2}\subset\mathbb{C}^{2}\times\mathbb{C}^{2}$.
But we are in $CP^{3}$, not $\mathbb{C}^{4}$, and the space described
by $\theta,\phi$ (or by $\xi,\varphi_{2}$) has only two dimensions
--- these two $S^{2}$ factors are not independent. In section \ref{sub:Subspace-S2xS2}
below we discuss a genuine four-dimensional $S^{2}\times S^{2}$ subspace.

The charges of this solution are very simply related to those of the
magnon on the sphere, since the extra term in the $CP^{3}$ angular
momentum \eqref{eq:Defn-Jz1-with-rho} compared to the that for the
sphere vanishes: $J_{\mathrm{sphere}}=J_{1}=\frac{1}{2}(J_{1}-J_{4})$,
and we get simply\begin{equation}
\Delta-\frac{J_{1}-J_{4}}{2}=2\sqrt{2\lambda}\:\sin\left(\frac{p}{2}\right).\label{eq:GHO-disp-rel}\end{equation}

One difference from the magnon on $S^{2}$ is that when $p=\pi$,
the magnon becomes a single closed string. Its cusps, at opposite
points on the equator of $S^{2}$, are in fact at the same point in
$RP^{2}$. In general the magnon connects two points a distance $\Delta\varphi_{2}=2\Delta\phi_{\mathrm{mag}}=2p$
apart on the equator, but $\varphi\sim\varphi+2\pi$ so $p=\delta$
and $p=\pi+\delta$ both connect the same two points. As was noted
by \cite{Gaiotto:2008cg}, this can be viewed as giving rise to a
second class of magnons, with \[
\Delta-\frac{J_{1}-J_{4}}{2}=2\sqrt{2\lambda}\:\sin\left(\frac{\pi+\delta}{2}\right)=2\sqrt{2\lambda}\:\cos\left(\frac{\delta}{2}\right).\]
Figure \ref{fig:Two-magnons} shows two magnons on $S^{2}$ and then
on $RP^{2}$, one with $p=\tfrac{1}{2}$ and another with $p=\pi-\tfrac{1}{2}$.
In the $RP^{2}$ case they have opposite opening angles $\delta=\pm\frac{1}{2}$,
thus form a single closed string, while in the $S^{2}$ case the total
opening angle is $\pi$.

\subsection{The subspace \rpthree\label{sub:Subspace-RP3}}

In the $AdS_{5}\times S^{5}$ case, Dorey's giant magnons with a second
large angular momentum $J'\sim\sqrt{\lambda}$ allow one to see that
the dispersion relation is $\Delta-J_{\mathrm{sphere}}=\sqrt{J^{\prime2}+\frac{\lambda}{\pi^{2}}\sin^{2}(p/2)}$.
\cite{Dorey:2006dq,Chen:2006gea} These necessarily live in $S^{3}$
rather than $S^{2}$. They are called dyonic magnons, and (embedding
$S^{3}\subset\mathbb{C}^{2}$) can be written\begin{align*}
W_{1} & =e^{it}\left(\cos\frac{p}{2}+i\sin\frac{p}{2}\,\tanh U\right)\\
W_{2} & =e^{iV}\sin\frac{p}{2}\,\sech U\end{align*}
where \begin{align*}
U & =\left(x\cosh\beta-t\sinh\beta\right)\cos\alpha & \cot\alpha & =\frac{2r}{1-r^{2}}\sin\frac{p}{2}\\
V & =\left(t\cosh\beta-x\sinh\beta\right)\sin\alpha & \tanh\beta & =\frac{2r}{1+r^{2}}\cos\frac{p}{2}.\end{align*}
The parameter $p$ is still the opening angle along the equator in
the $W_{1}$ plane, although $\cos(p/2)$ is clearly no longer the
worldsheet velocity. Sending the new parameter $r\to1$ reproduces
the original giant magnon.

The second method of embedding $S^{2}$ solutions into $CP^{3}$,
given by \eqref{eq:GHO-magnon-Z}, points out a way to embed $S^{3}$
solutions: \begin{equation}
\vec{Z}=\tfrac{1}{\sqrt{2}}\left(W_{1},W_{2},\bar{W}_{2},\bar{W}_{1}\right).\label{eq:RP3-embedding}\end{equation}
 As before, this is in fact a subspace $RP^{3}$ rather than $S^{3}$,
thanks to the identification of $(w_{1},w_{2})\sim-(w_{1},w_{2})$
implied.%
\footnote{Note that the rotation from $\vec{z}$ to $\vec{w}$ given by \eqref{eq:rotate45-defn-w}
is not an isometry, and in particular that the identification $\vec{z}\sim\lambda\vec{z}$
which defines $CP^{3}$ does not apply afterwards: $\vec{w}\nsim\lambda\vec{w}$
for complex $\lambda$. If $w_{3}=w_{4}=0$, as is implied by \eqref{eq:RP3-embedding},
then the phases of $w_{1}$ and $w_{2}$ are both physical. (Which
is good if we're claiming that the dyonic magnon has momenum along
both of them.)

However, the relation $\vec{w}\sim\lambda\vec{w}$ is true for real
$\lambda$, and since we have fixed $w_{1}^{2}+w_{2}^{2}=1$ by starting
with a string solution on $S^{2}$, the identification $(w_{1},w_{2})\sim-(w_{1},w_{2})$
is all that survives.%
}

Embedding a dyonic giant magnon in this way gives a $CP^{3}$ solution
with charges%
\footnote{In calculating these charges from \eqref{eq:Defn-Jz1-with-rho}, the
same cancellation of the second term happens here as happened in the
previous section. Thus using the charges one would expect for $S^{7}\subset\mathbb{C}^{4}$
gives the right answer here. This does not work in the $CP^{1}$ case,
see footnote \ref{fn:Note-on-wrong-J-defn}.%
} \begin{align*}
\Delta-\frac{J_{1}-J_{4}}{2} & =2\sqrt{2\lambda}\frac{1+r^{2}}{2r}\sin\left(\frac{p}{2}\right)\\
\frac{J_{2}-J_{3}}{2} & =2\sqrt{2\lambda}\frac{1-r^{2}}{2r}\sin\left(\frac{p}{2}\right).\end{align*}
These satisfy the relation \[
\Delta-\frac{J_{1}-J_{4}}{2}=\sqrt{\left(\frac{J_{2}-J_{3}}{2}\right)^{2}+8\lambda\sin^{2}\left(\frac{p}{2}\right)}.\]
Notice that the second angular momentum here is that carried by $Y^{2}$
and $Y_{3}^{\dagger}$, which are the impurities we insert into the
vacuum \eqref{eq:MZ-operator-vacuum} to make magnons in the $SU(2)\times SU(2)$
sector. 

This subspace can also be obtained from \eqref{eq:CP3-metric-Gibbons},
by fixing $\vartheta_{1}=\tfrac{\pi}{2}$, $\vartheta_{2}=\tfrac{\pi}{2}$
and $\eta=0$. The metric becomes \[
ds^{2}=d\xi^{2}+\sin^{2}\xi\: d\!\left(\frac{\varphi_{2}}{2}\right)^{2}+\cos^{2}\xi\: d\!\left(\frac{\varphi_{1}}{2}\right)^{2}.\]
This restriction can be checked to be a legal one from the equations
of motion for the angles $\vartheta_{1}$, $\vartheta_{2}$ and $\eta$.
The dyonic giant magnon in this space was re-derived by \cite{Ryang:2008rc},
using exactly these angles. It was also re-derived by \cite{Ahn:2008hj}
using co-ordinates $\vec{z}$.

Like the $RP^{2}$ magnons above, at $p=\pi$ these form single closed
strings, and beyond this ($\pi<p<2\pi$) give a second class of magnons
connecting the same two points on the equator as the magnon with $\tilde{p}=p-\pi$.

\section{Some larger subspaces\label{sec:Other-subspaces}}

All of the solutions we have discussed so far are known from the $AdS_{5}\times S^{5}$
case, and explore only subspaces $S^{2}$ or $S^{3}\subset S^{5}$.
In this section look at two subspaces of $CP^{3}$ on which new solutions
might exist: $CP^{2}$ and $S^{2}\times S^{2}$. 

We also study restrictions of this $S^{2}\times S^{2}$ down to three
or two dimensions (in sections \ref{sub:Subspace-S2-S1} and \ref{sub:S2-again})
since the resulting spaces have been used in the literature.

\subsection{The subspace \cptwo\label{sub:Subspace-CP2}}

The first larger nontrivial subspace we can find is $CP^{2}$, obtained
by setting $z_{3}=0$. In terms of the angles \eqref{eq:CP3-metric-Gibbons},
the restriction is $\vartheta_{2}=0$ (and $\varphi_{2}=0$, since
this is now redundant) and the metric becomes\[
ds^{2}=d\xi^{2}+\tfrac{1}{4}\cos^{2}\xi\left(d\vartheta_{1}^{2}+\sin^{2}\vartheta_{1}\: d\varphi_{1}^{2}\right)+\tfrac{1}{4}\sin^{2}2\xi\:\left(d\eta+\tfrac{1}{2}\cos\vartheta_{1}\, d\varphi_{1}\right)^{2}.\]
The two manifest isometries here are along $\varphi_{1}$ and $\eta$.
When $\xi=0$ this is an $S^{2}$ equivalent to \eqref{eq:S2-alpha-zero}
(exchange $z_{2}\leftrightarrow z_{4}$ to align them perfectly).
Perhaps allowing $\xi\neq0$ will allow new dyonic solutions here,
generalising the $CP^{1}$ solution \eqref{eq:GGY-magnon-Z} just
as the dyonic $RP^{3}$ solution generalises the $RP^{2}$ solution. 

Note that this is certainly a legal subspace, for the same reason
as given for $CP^{1}$: setting $z_{3}=0$ certainly solves the $z_{3}$
equation of motion.

\subsection{The subspace \ssfour \label{sub:Subspace-S2xS2}}

If we set $\varphi_{1}=\varphi_{2}$ and $\vartheta_{1}=\vartheta_{2}$
in metric \eqref{eq:CP3-metric-Gibbons}, we get the four-dimensional
space \begin{equation}
ds^{2}=\tfrac{1}{4}\left[d(2\xi)^{2}+\sin^{2}(2\xi)\: d\eta^{2}\right]+\tfrac{1}{4}\left[d\vartheta^{2}+\sin^{2}\vartheta\: d\varphi^{2}\right]\label{eq:four-dim-S2xS2}\end{equation}
which is $S^{2}\times S^{2}$ (possibly up to co-ordinate ranges),
and of course the new angles are defined $\vartheta\equiv(\vartheta_{1}+\vartheta_{2})/2$
and $\varphi\equiv(\varphi_{1}+\varphi_{2})/2$.

On such a product space, the Polyakov action splits into two terms,
giving two non-interacting sets of target-space co-ordinates. Any
two $S^{2}$ string solutions can be placed onto the same worldsheet,
completely independently. Choosing giant magnon solutions, worldsheet
scattering between these sectors would be trivial, just as it would
be on two decoupled Heisenberg spin chains. 

The restrictions needed to obtain this space are that $\vartheta_{-}\equiv\vartheta_{1}-\vartheta_{2}=0$
and $\varphi_{-}\equiv\varphi_{1}-\varphi_{2}=0$, and unfortunately
the equations of motion for $\vartheta_{-}$ and $\varphi_{-}$ are
not automatically solved by this choice: instead they give complicated
relations between the other co-ordinates. The equation for $\vartheta_{-}$
reads\begin{align*}
0 & =-\partial_{t}\left(\cos2\xi\,\partial_{t}\vartheta\right)+\partial_{x}\left(\cos2\xi\,\partial_{x}\vartheta\right)+\tfrac{1}{2}\cos2\xi\sin2\vartheta\left(\partial_{t}^{2}\varphi-\partial_{x}^{2}\varphi\right)\\
 & \qquad-\sin^{2}2\xi\sin\vartheta\left(\partial_{t}\eta\,\partial_{t}\varphi-\partial_{x}\eta\,\partial_{x}\varphi\right)\end{align*}
and that for $\varphi_{-}$ reads\begin{align*}
0 & =-\partial_{t}\left(\sin^{2}2\xi\cos\vartheta\,\partial_{t}\eta+\cos2\xi\sin^{2}\vartheta\,\partial_{t}\varphi\right)\\
 & \qquad+\partial_{x}\left(\sin^{2}2\xi\cos\vartheta\,\partial_{x}\eta+\cos2\xi\sin^{2}\vartheta\,\partial_{x}\varphi\right).\end{align*}
These constraints do not of course rule out the existence of solutions
on this subspace. But placing an arbitrary $S^{2}$ solution into
each of the factors is unlikely to produce a solution, because of
these equations coupling $\xi,\eta$ to $\vartheta,\varphi$.

\subsection{The subspace \ssthree\label{sub:Subspace-S2-S1}}

If we further restrict the above subspace by holding one of the angles
fixed, we will get $S^{2}\times S^{1}$ (again up to identifications).
Setting $\vartheta=\frac{\pi}{2}$ gives the space studied by \cite{Ryang:2008rc},
with metric\[
ds^{2}=\tfrac{1}{4}\left[d(2\xi)^{2}+\sin^{2}(2\xi)\, d\eta^{2}+d\varphi^{2}\right].\]
The equation of motion for $\vartheta$ is solved by $\vartheta=\frac{\pi}{2}$,
and the constraints imposed by $\vartheta_{-}=0$ and $\varphi_{-}=0$
above simplify to \begin{equation}
0=-\partial_{t}\eta\:\partial_{t}\varphi+\partial_{x}\eta\:\partial_{x}\varphi\qquad\qquad\qquad\label{eq:First-constr-S2xS1}\end{equation}
\begin{equation}
0=-\partial_{t}\left(\cos2\xi\:\partial_{t}\varphi\right)+\partial_{x}\left(\cos2\xi\:\partial_{x}\varphi\right).\label{eq:Second-constr-S2xS1}\end{equation}

These constraints were not taken into account by \cite{Ryang:2008rc},
who sets $\vartheta_{-}=0$ before calculating the equation of motion
for $\vartheta$ (which is indeed solved) but without ever calculating
the equation of motion for $\vartheta_{-}$.%
\footnote{The constraint \eqref{eq:First-constr-S2xS1} can also be obtained
without using $\vartheta_{-}$, by simply setting $\vartheta_{1}=\frac{\pi}{2}$
and $\vartheta_{2}=\frac{\pi}{2}$ in their equations of motion.%
} The magnon ansatz used there sets $\eta=\omega t+f(u)$, $\varphi=\nu t$
and $\xi=g(u)$, in terms of boosted $u=\beta t+\alpha x$. The first
constraint \eqref{eq:First-constr-S2xS1} then implies $\beta\, f'(u)=-\omega$,
while for a magnon solution one typically has $f(u)\propto\tanh u$.
The second constraint \eqref{eq:Second-constr-S2xS1} implies $\beta=0$,
so together they imply $\omega=0$. 

This problem does not arise in the other case studied by \cite{Ryang:2008rc},
where the $\vartheta_{-}$ equation is solved by $\eta=0$, and $\varphi_{1}\neq\varphi_{2}$
so there is no $\varphi_{-}$ constraint. The resulting subspace is
the $RP^{3}$ discussed in section \ref{sub:Subspace-RP3}.

\subsection{The subspace \cpone , again\label{sub:S2-again}}

Finally, we can restrict the subspace $S^{2}\times S^{2}$ of \eqref{eq:four-dim-S2xS2}
by holding both of the angles in one factor constant, to obtain $S^{2}$.
Setting $\xi$ and $\eta$ to be constants leaves the space \[
ds^{2}=\tfrac{1}{4}\left[d\vartheta^{2}+\sin^{2}\vartheta\: d\varphi^{2}\right]\]
which is, like our $CP^{1}$ of section \ref{sub:Subspace-CP1}, a
sphere of radius $\frac{1}{2}$. This is a legal subspace, as the
equations of motion for $\xi$ and $\eta$ are automatically solved
(because a stationary particle anywhere on the sphere is a solution)
and the constraints arising from $\vartheta_{-}=0$ and from $\varphi_{-}=0$
become simply the equations of motion for $\vartheta$ and $\varphi$.

When $\xi=\frac{\pi}{2}$, and using the conventions given in appendix
\ref{sec:more-metric-details}, this space is embedded by\[
\vec{z}=\left(e^{i\varphi/2}\cos\frac{\vartheta}{2}\,,\:0,\:0,\: e^{-i\varphi/2}\sin\frac{\vartheta}{2}\right).\]
This is precisely the same subspace $CP^{1}$ as in \eqref{eq:S2-alpha-zero},
although we obtained it there by fixing $\alpha=0$ in the other set
of angles \eqref{eq:CP3-metric-Pope}. Fixing $\xi$ to some other
value will simply rotate the 1-2 and 3-4 planes, but in all cases
the space is $S^{2}=CP^{1}$. Like the subspace $RP^{2}$ discussed
in section \ref{sub:Subspace-RP2}, this one is sometimes referred
to as $S^{2}\times S^{2}$ in the literature.

These co-ordinates were used by \cite{Lee:2008ui} to study finite-$J$
effects on the $CP^{1}$ giant magnon. We give their results in \eqref{eq:finite-J-disp-CP1}
below.

\section{Finite-\emph{J} corrections\label{sec:Finite-J-corrections}}

All of the giant magnons we have written down so far have both infinite
energy and infinite angular momentum. As can be seen from \eqref{eq:defn-Delta-timelike-gauge},
this corresponds to infinite worldsheet length in the timelike conformal
gauge we are using. 

The first treatment of giant magnons $AdS_{5}\times S^{5}$ at finite
$J$ was by \cite{Arutyunov:2006gs}, who worked in uniform lightcone
gauge, in which the worldsheet density of $J$, rather than of $\Delta$,
is constant. Their gauge has a parameter $a\in[0,1]$, and at $a=0$
(and in conformal gauge) they obtained the following correction to
the dispersion relation: \begin{align*}
\varepsilon\equiv\Delta-J & =\frac{\sqrt{\lambda}}{\pi}\sin\left(\frac{p}{2}\right)\left[1-\frac{4}{e^{2}}\sin^{2}\left(\frac{p}{2}\right)e^{-2J/\varepsilon}+o(e^{-4J/\varepsilon})\right]\\
 & =\frac{\sqrt{\lambda}}{\pi}\sin\left(\frac{p}{2}\right)\left[1-4\sin^{2}\left(\frac{p}{2}\right)\: e^{-2\Delta/\varepsilon}+\ldots\right]\end{align*}
Exact solutions at any $J$ were studied by \cite{Okamura:2006zv},
where it was shown that they are connected by the Pohlmeyer map to
kink-train solutions of sine-gordon theory. The apparent gauge-dependence
of the results of \cite{Arutyunov:2006gs} was resolved by \cite{Astolfi:2007uz},
using the fact that the solutions are periodic both on the worldsheet
and in the azimuthal angle on the sphere, and so can be viewed as
wound strings on $S^{2}/\mathbb{Z}_{n}$. \cite{Astolfi:2007uz,Ramadanovic:2008qd}
The scattering of finite-$J$ magnons was studied by \cite{Klose:2008rx},
using the connection to sine-gordon theory in finite volume.

The finite-$J$ generalisations of the basic giant magnon are still
solutions moving on $S^{2}$, and so one can place them into $CP^{3}$
using either of the maps presented in sections \ref{sub:Subspace-CP1}
and \ref{sub:Subspace-RP2} above. For the $RP^{2}$ giant magnon,
the corrected dispersion relation was derived by \cite{Grignani:2008te}
to be\begin{equation}
\Delta-\frac{J_{1}-J_{4}}{2}=2\sqrt{2\lambda}\sin\left(\frac{p}{2}\right)\left[1-4\sin^{2}\left(\frac{p}{2}\right)e^{-2\Delta\big/2\sqrt{2\lambda}\sin(\frac{p}{2})}+\ldots\right].\label{eq:finite-J-disp-RP2}\end{equation}
For the $CP^{1}$ giant magnon, \cite{Lee:2008ui} give the result%
\footnote{Here is brief note about deriving these two results from the original
 $S^{2}$ case. The integrals defining the charges are now over a
finite length $-L<x<L$, so write $J(L)$ and $\Delta(L)$. Note that
$\Delta(2L)=2\Delta(L)$. To get the charges for one magnon, we must
integrate from one cusp to the next: choose $L$ such that $\theta_{\mathrm{mag}}(x=\pm L,t=0)$
are at the first cusps. 

For the $RP^{2}$ case, the relationship we used before $J_{\mathrm{sphere}}(L)=J_{1}(L)=\left(J_{1}(L)-J_{4}(L)\right)/2$
still holds, leading to \eqref{eq:finite-J-disp-RP2}. We wrote the
$S^{2}$ result above using the prefactor appropriate for $AdS_{5}\times S^{5}$,
so to get this result for the $AdS_{4}\times CP^{3}$ theory have
replaced $\sqrt{\lambda}/\pi\to2\sqrt{2\lambda}$.

For the $CP^{1}$ case, the cusp at $\theta_{\mathrm{mag}}(L,0)$
is at $\vec{Z}_{CP^{1}}(\frac{L}{2},0)$, thanks to the scaling \eqref{eq:2x-2t-scaling}.
The relationship between charges is that \[
\frac{J_{1}(\frac{L}{2})-J_{4}(\frac{L}{2})}{2}=\frac{1}{2}\, J_{\mathrm{sphere}}(L).\]
Thus $\Delta(\frac{L}{2})-(J_{1}(\frac{L}{2})-J_{4}(\frac{L}{2}))/2=\Delta(\frac{L}{2})-\frac{1}{2}J_{\mathrm{sphere}}(L)=\frac{1}{2}\left(\Delta(L)-J_{\mathrm{sphere}}(L)\right)$.
In the result \eqref{eq:finite-J-disp-CP1}, it is the energy for
one magnon $\Delta(\frac{L}{2})$ which appears both on the left hand
side and in the exponent. %
} \begin{equation}
\Delta-\frac{J_{1}-J_{4}}{2}=\sqrt{2\lambda}\sin\left(\frac{p}{2}\right)\left[1-4\sin^{2}\left(\frac{p}{2}\right)e^{-2\Delta\big/\sqrt{2\lambda}\sin(\frac{p}{2})}+\ldots\right].\label{eq:finite-J-disp-CP1}\end{equation}
We observe that, even at finite $J$, two $CP^{1}$ magnons have the
same dispersion relation as one $RP^{2}$ magnon, provided all three
have the same value of the parameter $p$.%
\footnote{Note that that essentially all the properties of the two $CP^{1}$
magnons add up to give those of the single $RP^{2}$ magnon: energy
$\Delta$, angular momentum $(J_{1}-J_{4})/2$, worldsheet length
$L$ and opening angle along the equator (which we call $\Delta\beta$
in the next section).%
} 

Dyonic giant magnons can also be studied at finite $J$; this has
been done for those in $S^{5}$ from this string sigma-model perspective
by \cite{Okamura:2006zv,Hatsuda:2008gd}, and for those in $RP^{3}\subset CP^{3}$
by \cite{Ahn:2008hj,Ahn:2008wd}.

In the $AdS_{5}\times S^{5}$ case these corrections can also be calculated
using algebraic curves \cite{Kazakov:2004qf,Beisert:2004ag,SchaferNameki:2004ik,Beisert:2005bm,Minahan:2006bd,Vicedo:2007rp,Gromov:2007aq,Minahan:2008re,Gromov:2008ec,Sax:2008in}
or using the L\"{u}scher formula \cite{Luscher:1985dn,Janik:2007wt,Gromov:2008ie,Heller:2008at,Hatsuda:2008na},
and these agree with the string sigma-model result presented above.
For calculations on the gauge theory side of the correspondance see
\cite{Serban:2004jf,Sieg:2005kd,Ambjorn:2005wa,SchaferNameki:2006ey,Kotikov:2007cy,Fiamberti:2007rj,Keeler:2008ce,Fiamberti:2008sh}.
In $AdS_{4}\times CP^{3}$ the same list of methods is possible, and
we discuss these further in section \ref{sub:Beyond-the-classical}
below.

\section{Discussion and conclusion\label{sec:Discussion-and-conclusion}}

In this paper we have only discussed giant magnon solutions known
from $AdS_{5}\times S^{5}$, but have been careful about how these
are placed into $CP^{3}$. Here we summarise these results, comment
on more general solutions, and comment on connections to approaches
other than the classical string sigma-model.

\subsection{Single-charge giant magnons}

In sections \ref{sub:Subspace-CP1} and \ref{sub:Subspace-RP2} we
looked at two different ways to embed the basic single-charge giant
magnon \eqref{eq:basic-magnon}, into either $CP^{1}$ or $RP^{2}$.
\cite{Gaiotto:2008cg,Grignani:2008is} This $CP^{1}$ is a two-sphere
of radius $\frac{1}{2}$, while $RP^{2}$ is half a two-sphere, so
both have an equator of length $\pi$. We lined up the embeddings
into $\mathbb{C}^{4}$ such that, in both cases, the equator is the
line \[
\vec{z}=\tfrac{1}{\sqrt{2}}\left(e^{i\beta},0,0,e^{-i\beta}\right)\]
where we name the angle $\beta\in[0,\pi]$, as in \eqref{eq:strange-new-co-ords}
above, to avoid confusion.

Since the basic magnon \eqref{eq:basic-magnon} has opening angle
$\Delta\phi_{\mathrm{mag}}=p$, these two solutions have\begin{align*}
CP^{1}:\qquad\beta & =\chi/4=\phi_{\mathrm{mag}}/2\qquad\implies\quad\Delta\beta=p/2\\
RP^{2}:\qquad\beta & =\varphi_{2}/2=\phi_{\mathrm{mag}}\;\:\qquad\implies\quad\Delta\beta=p'\end{align*}
(where we now write $p'$ for the parameter of the $RP^{2}$ magnon,
to distinguish it from the $CP^{1}$ case's $p$). A single giant
magnon is not a closed string solution, one must join a set of them
together at their endpoints on the equator. The condition for a set
$p_{i}$ of $CP^{1}$ magnons or $p'_{j}$ of $RP^{2}$ magnons to
close is that the total opening angle $\Delta\beta$ should be a multiple
of $\pi$, that is, \begin{align}
CP^{1}:\:\;\qquad\sum_{i}p_{i} & =2\pi n\label{eq:colsure-condition-p}\\
RP^{2}:\qquad\sum_{j}2p'_{j} & =2\pi n\,,\qquad n\in\mathbb{Z}.\nonumber \end{align}

The point particle \eqref{eq:point-particle} moves along the same
equator too, and by calculating fluctuations of this solution, we
checked in section \ref{sec:Fluctuation-Hamiltonian} that $\Delta-\frac{J_{1}-J_{4}}{2}$
is indeed a Hamiltonian for them, just as $\Delta-J$ is in the $S^{5}$
case. Calculating the same difference of charges for the two magnon
embeddings, we obtained dispersion relations \eqref{eq:GGY-disp-rel}
and \eqref{eq:GHO-disp-rel}, which we now write also in terms of
the opening angle $\Delta\beta$: \begin{align*}
CP^{1}:\qquad\Delta-\frac{J_{1}-J_{4}}{2} & =\sqrt{2\lambda}\sin\left(\frac{p}{2}\right)\quad\:=\sqrt{2\lambda}\sin\left(\Delta\beta\right)\\
RP^{2}:\qquad\Delta-\frac{J_{1}-J_{4}}{2} & =2\sqrt{2\lambda}\sin\left(\frac{p'}{2}\right)=2\sqrt{2\lambda}\sin\left(\frac{\Delta\beta}{2}\right).\end{align*}
Notice that these agree at small $\Delta\beta$. The limit $p\to0$
takes you from giant magnons to the Penrose limit (via the interpolating
case of \cite{Maldacena:2006rv}, studied here by \cite{Kreuzer:2008vd}).
Finite-$J$ effects in the Penrose limit were studied by \cite{Astolfi:2008ji}. 

As noted in section \ref{sub:Subspace-RP2}, there is also a second
magnon on $RP^{2}$ for any given opening angle $\Delta\beta$, which
has charges \cite{Gaiotto:2008cg} \begin{align*}
RP^{2\prime}:\qquad\Delta-\frac{J_{1}-J_{4}}{2} & =2\sqrt{2\lambda}\cos\left(\frac{\Delta\beta}{2}\right).\qquad\qquad\end{align*}
For small $\Delta\beta$ this is almost a circular string, with its
ends slightly offset along the equator --- see figure \ref{fig:Two-magnons}
on page \pageref{fig:Two-magnons} above.

\subsection{More solutions! \label{sub:More-solutions!}}

While we used the giant magnon on $S^{2}$ \eqref{eq:basic-magnon}
as an example, the subspaces we have described exist independently
of it, and any other string solution moving on $S^{2}$ can be placed
into either of these subspaces of $CP^{3}$ in the same way. Thus
not only finite-$J$ magnons (as discussed in section \ref{sec:Finite-J-corrections}
above) but also scattering solutions \cite{Spradlin:2006wk,Chen:2006gq,Kalousios:2008gz}
and single spikes%
\footnote{Single-spike solutions of all kinds can be easily obtained from their
giant magnon partners by the $x\leftrightarrow t$ exchange discussed
in \cite{Ishizeki:2007we,Abbott:2008yp}. As in $\mathbb{R}\times S^{5}$,
this exchange (keeping $X^{0}=t$) is a symmetry of the equations
of motion \eqref{eq:eq-of-m-in-Z} and the Virasoro constraints for
$\mathbb{R}\times CP^{3}$. Thus the classical solutions have no properties
which cannot be read off from the corresponding magnon solution. However,
the quantum properties are quite different. \cite{Abbott:2008yp}%
} \cite{Ishizeki:2007we,Abbott:2008yp,Mosaffa:2007ty,Hayashi:2007bq,Ahn:2008sk,Jain:2008mt}
all exist in both the $CP^{1}$ and $RP^{2}$ subspaces. The equations
of motion do not notice the global identification $(w_{1},w_{2})\sim-(w_{1},w_{2})$
which distinguishes $RP^{2}$ from $S^{2}$, and the fact that $CP^{1}$
is a sphere of radius $\frac{1}{2}$ can be dealt with by the same
scaling \eqref{eq:2x-2t-scaling} that we used for the basic magnon.

Many papers interpret the magnon on $RP^{2}$ (and also that on $RP^{3}$)
as being two magnons, one in each half of the embedding space $\mathbb{C}^{2}\times\mathbb{C}^{2}$.
\cite{Grignani:2008is,Ahn:2008hj} It is then tempting to identify
these two halves with the even- and odd-site spin chains in the dual
description's $SU(2)\times SU(2)$ sector. For the known solutions,
however, these two halves are not independent: in fact they are always
locked together, and by a trivial change of co-ordinates \eqref{eq:rotate45-defn-w}
we can write them as a single $RP^{2}=S^{2}/\mathbb{Z}_{2}$ space.
This does not rule out the existence of two independent magnon sectors,
such that a pair of magnons of the same parameter $p$ gives us again
the known $RP^{2}$ solution. But at present individual solutions
in these two sectors are not known.

The single-parameter giant magnon on $S^{2}$ has a two-parameter
dyonic generalisation on $S^{3}$, and in section \ref{sub:Subspace-RP3}
we looked at how to map this into $RP^{3}\subset CP^{3}$, where it
generalises the $RP^{2}$ solution. The dyonic generalisation of the
$CP^{1}$ solution is not known, but it might lie in the $CP^{2}$
subspace we discussed in section \ref{sub:Subspace-CP2}. 

It would be very interesting to find some indication among the magnon
solutions of the weaker momentum constraint: the momentum in just
the even-site or just the odd-site spin chain need not vanish, only
the total. Combining the two closure conditions \eqref{eq:colsure-condition-p}
to give $\sum_{i}p_{i}+\sum_{j}2p_{j}'=2\pi n$ cannot be the answer,
because these two classes of magnons are certainly inequivalent solutions,
while the even- and odd-site spin chains are related by an $SU(4)$
rotation.

\subsection{Beyond the classical sigma-model\label{sub:Beyond-the-classical}}

The classical string solutions we have discussed are well-known from
the $S^{5}$ case, and explore only $S^{2}$ or $S^{3}$-like subspaces
of $CP^{3.}$. Their classical properties (and indeed those of solutions
we have not discussed, such as scattering solutions) are not strongly
affected by being transplanted to the new space. However, their quantum
properties will certainly depend on the whole space, as was the case
for spinning string solutions in $AdS_{2}\times S^{1}$ studied by
\cite{McLoughlin:2008ms,Alday:2008ut,Krishnan:2008zs}. The relevant
supersymmetric sigma model (for strings on $AdS_{4}\times CP^{3}$)
was first studied by \cite{Arutyunov:2008if,Stefanski:2008ik}. Using
this one would like to perform a calculation like that done for magnons
in $AdS_{5}\times S^{5}$ by \cite{Minahan:2007gf,Papathanasiou:2007gd}. 

Like the equations of motion, the Pohlmeyer map \cite{Pohlmeyer:1975nb,Lund:1976ze}
to the sine-gordon field $\alpha$ (given by $\cos\alpha=-\partial_{t}\bar{W}_{i}\partial_{t}W_{i}+\partial_{x}\bar{W}_{i}\partial_{x}W_{i}$
in the $S^{2}$ case) depends only locally on the target-space co-ordinates.
Thus strings on either $CP^{1}$ or $RP^{2}$ will be classically
equivalent to the sine-gordon model. The condition that the string
closes $\sum\Delta\beta\sim0$ plays no role in the sine-gordon model,
thus the second class of magnons, which we called $RP^{2\prime}$
above, has no special meaning in sine-gordon theory. As quantum systems,
strings on $\mathbb{R}\times S^{2}$ are quite different to the sine-gordon
model, thanks to the different notion of energy, and this complicates
the translation of the $n$-body description of solitons in sine-gordon
theory to this case. \cite{Mikhailov:2005sy,Hofman:2006xt,Ruijsenaars:1986vq,Babelon:1993bx,Aniceto:2008pc}
The Pohlmeyer reduction has been extended to the full superstring
on $AdS_{5}\times S^{5}$, \cite{Grigoriev:2007bu,Mikhailov:2007xr,Grigoriev:2008jq}
and also to strings moving on $CP^{3}$. \cite{Eichenherr:1979uk,Rashkov:2008rm,Miramontes:2008wt}.

Classical strings in $AdS_{4}\times CP^{3}$ can also be studied using
the algebraic curve, in which the 10 eigenvalues $q_{a}$ of the monodromy
matrix $\Omega$ are analytic functions of the spectral parameter,
and their various poles and branch points control the solution. \cite{Gromov:2008bz}
Giant magnons in this picture were studied by \cite{Shenderovich:2008bs},
and are of two distinct kinds, `small' and `big'. Their dispersion
relations are as follows:\begin{align*}
\mbox{small GM:}\qquad\qquad\varepsilon & =\sqrt{\frac{1}{4}+2\lambda\sin^{2}\left(\frac{p}{2}\right)}\quad\to\sqrt{2\lambda}\:\sin\left(\frac{p}{2}\right)\quad\mbox{ when }\sqrt{\lambda}\gg1\\
\mbox{big GM:}\qquad\qquad\varepsilon & =\sqrt{1+8\lambda\sin^{2}\left(\frac{p}{4}\right)}\quad\to2\sqrt{2\lambda}\:\sin\left(\frac{p}{4}\right).\end{align*}
It would seem natural to identify these with the $CP^{1}$ and $RP^{2}$
magnons of the string sigma-model, presumably with $p'=p/2=\Delta\beta$.
There are two `small GM' sectors, together often called the $SU(2)\times SU(2)$
sector. 

However, the study of finite-$J$ corrections to these paints a different
picture. According to \cite{Lukowski:2008eq}, two `small GM's in
the two sectors, both with the same momentum $p$, have a correction
$\delta\varepsilon$ matching the $RP^{2}$ string result \eqref{eq:finite-J-disp-RP2}.
This does seems to point to the interpretation of the $RP^{2}$ string
solution as two giant magnons, as was originally claimed by \cite{Grignani:2008is}.
However, the same paper's result for one `small GM' does not match
any of the string calculations, apparently leaving open the identification
both of the string state for this, and of the algebraic curve corresponding
to the $CP^{1}$ string. Finite-$J$ corrections have also been studied
using the L\"{u}scher formula by \cite{Bombardelli:2008qd,Lukowski:2008eq},
and the results agree with those from the algebraic curve.

\section*{Acknowledgements}

We would like to thank Antal Jevicki and Marcus Spradlin for helpful
comments on a draft of this paper, and Olof Ohlsson Sax for correspondence
about finite-$J$ effects. 

This work was supported in part by DOE grant DE-FG02-91ER40688-Task~A.
IA was also supported in part by POCI 2010 and FSE, Portugal, through
the fellowship SFRH/BD/14351/2003. MCA would also like to thank the
Mathematics Department for financial support.

\bigskip \bigskip \bigskip 

\appendix

\section{More about \cpthree's geometry \label{sec:more-metric-details}}

The complex projective space $CP^{3}$ is defined to be \[
CP^{3}=\frac{\mathbb{C}^{4}}{\vec{z}\sim\lambda\vec{z}}\]
where $\vec{z}=z_{a}$ are called homogeneous co-ordinates. We can
split this identification into $\vec{z}\sim r\vec{z}$ and $\vec{z}\sim e^{i\phi}\vec{z}$
(for any $r,\phi\in\mathbb{R}$) and then replace the first one with
the condition $\left|\vec{z}\right|^{2}=1$, to obtain a sphere with
one identification\[
CP^{3}=\frac{S^{7}}{\vec{z}\sim e^{i\phi}\vec{z}}=\frac{S^{7}}{U(1)}\,.\]
The isometry group is $SU(4)$, acting in the natural way on $\vec{z}$.
Since the stabiliser group of (say) the point $z_{4}=1$ is $U(3)$,
we can also write\[
CP^{3}=\frac{SU(4)}{U(3)}\,.\]

The infinitesimal form of the standard Fubini--Study metric for this
is \begin{align}
ds_{CP^{3}}^{2} & =\frac{dz_{i}d\bar{z}_{i}}{\rho^{2}}-\frac{\left|z_{i}d\bar{z}_{i}\right|^{2}}{\rho^{4}}\nonumber \\
 & =ds_{\mathrm{sphere}}^{2}-d\gamma^{2}\label{eq:metric-sphere-minus-dgamma}\\
 & =\frac{ds_{\mathrm{flat}}^{2}-d\rho^{2}}{\rho^{2}}-d\gamma^{2}\nonumber \end{align}
where $\rho^{2}=z_{i}\bar{z}_{i}$. (Note that in some conventions
the metric is 4 times this, \cite{Kobayashi1963,Gaiotto:2008cg} making
$CP^{1}$ \eqref{eq:S2-alpha-zero} a unit sphere.) In the second
and third lines above, $ds_{\mathrm{flat}}^{2}=dz_{i}d\bar{z}_{i}$
is the Euclidean metric for $\mathbb{C}^{4}$, and $ds_{\mathrm{sphere}}^{2}$
is a metric for $S^{7}$ in terms of these embedding co-ordinates.
Instead of fixing $\rho=1$, this way of treating the sphere subtracts
off the component coming from radial motion (and scales the rest appropriately).
In turn, $CP^{3}$ can be obtained from the sphere by fixing the total
phase $\gamma=\mbox{arg}\prod_{i}z_{i}$, or instead by subtracting
the total phase component. These two pieces are\begin{align*}
d\rho & =\frac{1}{2\rho}\left(z_{i}d\bar{z}_{i}+\bar{z}_{i}dz_{i}\right)=\frac{1}{\rho}\re\left(\bar{z}_{i}dz_{i}\right)\\
d\gamma & =\frac{i}{2\rho^{2}}\left(z_{i}d\bar{z}_{i}-\bar{z}_{i}dz_{i}\right)=\frac{1}{\rho^{2}}\im\left(\bar{z}_{i}dz_{i}\right).\end{align*}

We now present the maps between the homogeneous co-ordinates and the
two sets of angles we have used. These are taken from \cite{Nishioka:2008gz}
and \cite{Pope:1984bd}, although we have shuffled the $z_{i}$. For
the metric \eqref{eq:CP3-metric-Gibbons} (whose $\eta$ is often
called $\psi$) \begin{align*}
ds_{CP^{3}}^{2} & =d\xi^{2}+\frac{1}{4}\sin^{2}2\xi\left(d\eta+\frac{1}{2}\cos\vartheta_{1}\: d\varphi_{1}-\frac{1}{2}\cos\vartheta_{2}\: d\varphi_{2}\right)^{2}\\
 & \qquad+\frac{1}{4}\cos^{2}\xi\left(d\vartheta_{1}^{2}+\sin^{2}\vartheta_{1}\: d\varphi_{1}^{2}\right)+\frac{1}{4}\sin^{2}\xi\left(d\vartheta_{2}^{2}+\sin^{2}\vartheta_{2}\: d\varphi_{2}^{2}\right)\end{align*}
the relationship is: \begin{align}
z_{1} & =\sin\xi\:\cos(\vartheta_{2}/2)\: e^{-i\eta/2}\: e^{i\varphi_{2}/2}\nonumber \\
z_{2} & =\cos\xi\:\cos(\vartheta_{1}/2)\: e^{i\eta/2}\: e^{i\varphi_{1}/2}\label{eq:map-Gibbons-to-z}\\
z_{3} & =\cos\xi\:\sin(\vartheta_{1}/2)\: e^{i\eta/2}\: e^{-i\varphi_{1}/2}\nonumber \\
z_{4} & =\sin\xi\:\sin(\vartheta_{2}/2)\: e^{-i\eta/2}\: e^{-i\varphi_{2}/2}.\nonumber \end{align}
For the other set of angular variables \eqref{eq:CP3-metric-Pope}
\begin{align*}
ds_{CP^{3}}^{2} & =d\mu^{2}+\frac{1}{4}\sin^{2}\mu\cos^{2}\mu\left[d\chi+\sin^{2}\alpha\:\left(d\psi+\cos\theta\; d\phi\right)\right]^{2}\\
 & \qquad+\sin^{2}\mu\left[d\alpha^{2}+\frac{1}{4}\sin^{2}\alpha\left(d\theta^{2}+\sin^{2}\theta\; d\phi^{2}+\cos^{2}\alpha\left(d\psi+\cos\theta\; d\phi\right)^{2}\right)\right]\end{align*}
the map is specified by \begin{align}
z_{1}/z_{4} & =\tan\mu\;\cos\alpha\; e^{i\chi/2}\nonumber \\
z_{2}/z_{4} & =\tan\mu\;\sin\alpha\;\sin(\theta/2)\; e^{i\chi/2}\; e^{i(\psi-\phi)/2}\label{eq:map-Pope-to-z}\\
z_{3}/z_{4} & =\tan\mu\;\cos\alpha\;\cos(\theta/2)\; e^{i\chi/2}\; e^{i(\psi+\phi)/2}.\nonumber \end{align}
These ratios $z_{i}/z_{4}$ are called inhomogeneous co-ordinates,
and cover the patch $z_{4}\neq0$ with no identifications. \cite{Kobayashi1963}
With the ranges given, the trigonometric functions controlling the
amplitudes are always positive in both of these cases. From the phases
of the inhomogeneous co-ordinates of $z_{i}/z_{4}$ it is easy to
see that ranges of the remaining angles are correct.

\section{Strings in homogeneous co-ordinates\label{sec:strings-sans-trig}}

To study bosonic string theory in $S^{n}$, it is often convenient
to use embedding co-ordinates for $\mathbb{R}^{n+1}$ and then constrain
the radius to 1. This avoids all the trigonometric functions needed
for angular co-ordinates, and (in AdS/CFT) also gives a simple correspondence
between the R-symmetry generators and the rotations of this space.
We can do the same for $CP^{3}$, using homogeneous co-ordinates $\vec{z}$.
We will need two constraints, $\rho^{2}=1$ and $\gamma=0$.

\subsection{Using Lagrange multipliers}

Begin by writing the metric for $\mathbb{R}\times CP^{3}$ as \[
ds^{2}=-\left(dX^{0}\right)^{2}+d\bar{z}_{i}G_{ij}dz_{j}\qquad\mbox{with }\; G_{ij}=\frac{\delta_{ij}}{\rho^{2}}-\frac{z_{i}\bar{z}_{j}}{\rho^{4}}\]
In conformal gauge, and with $X^{0}=\kappa t$, the Polyakov action
is\begin{align}
S & =\int\frac{dx\, dt}{2\pi}\, R^{2}\mathcal{L}\label{eq:Polyakov-action}\\
 & =2\sqrt{2\lambda}\int dx\, dt\,\mathcal{L}\nonumber \\
2\mathcal{L} & =\kappa^{2}+\partial^{a}\bar{Z}_{i}G_{ij}\partial_{a}Z_{j}+\Lambda_{\rho}\left(\bar{Z}_{i}Z_{i}-1\right)+i\Lambda_{\gamma}\left(Z_{1}Z_{2}Z_{3}Z_{4}-\bar{Z}_{1}\bar{Z}_{2}\bar{Z}_{3}\bar{Z}_{4}\right).\nonumber \end{align}
Note that $\Lambda_{\gamma}\in\mathbb{R}$, since the piece in brackets
is proportional to $2i\sin\gamma$. In calculating Euler--Lagrange
equations for this, we set $\rho=1$ immediately, simplifying $\partial G_{ij}/\partial Z_{i}$
etc. greatly. The Lagrange multipliers can be read off from the parallel
component of the equations (i.e. $\bar{Z}_{i}$ times $Z_{i}$'s equation
of motion) which is:\begin{align*}
\Lambda_{\rho}-4i\left(Z_{1}Z_{2}Z_{3}Z_{4}\right)\Lambda_{\gamma} & =\partial_{t}\bar{Z}_{i}\partial_{t}Z_{i}-2\left|\bar{Z}_{i}\partial_{t}Z_{i}\right|^{2}\;-\:\partial_{x}\bar{Z}_{i}\partial_{x}Z_{i}+2\left|\bar{Z}_{i}\partial_{x}Z_{i}\right|^{2}.\end{align*}
(This 4 is the number of complex embedding co-ordinates.) The right-hand
side here is real, which implies $\Lambda_{\gamma}=0$. Using this,
we find the equation of motion for $Z_{i}$ to be\begin{equation}
-\partial_{t}\left(G_{ij}\partial_{t}Z_{j}\right)+\partial_{x}\left(G_{ij}\partial_{x}Z_{j}\right)=Z_{i}\Lambda_{\rho}-\left(\bar{Z}_{j}\partial_{t}Z_{j}\right)\partial_{t}Z_{i}+\left(\bar{Z}_{j}\partial_{x}Z_{j}\right)\partial_{x}Z_{i}\:.\label{eq:eq-of-m-in-Z}\end{equation}
The Virasoro constraints are\[
-\kappa^{2}+\partial_{t}\bar{Z}_{i}\: G_{ij}\:\partial_{t}Z_{j}+\partial_{x}\bar{Z}_{i}\: G_{ij}\:\partial_{x}Z_{j}=0\]
\[
\re\left(\partial_{t}\bar{Z}_{i}\: G_{ij}\:\partial_{x}Z_{j}\right)=0\,.\]

The result that $\Lambda_{\gamma}=0$ deserves a little explanation.
If we were to analyse strings on the sphere using a similar metric
(in fact exactly $ds_{\mathrm{sphere}}^{2}$ from \eqref{eq:metric-sphere-minus-dgamma}
above): \[
2\mathcal{L}=1+\partial^{a}X_{i}\partial_{a}X_{j}g_{ij}+\Lambda(X^{2}-1),\qquad\mbox{with }\; g_{ij}=\frac{\delta_{ij}}{\rho^{2}}-\frac{X_{i}X_{j}}{\rho^{4}}\]
then we would also find $\Lambda=0$, although the equations of motion
are the same as are obtained with $g_{ij}=\delta_{ij}$ (i.e. using
$ds_{\mathrm{flat}}^{2}$). In some sense the metric is enforcing
the constraint for us. The reason we had $\Lambda_{\rho}\neq0$ in
the $CP^{3}$ case above was that we set $\rho=1$ at an early stage
of the calculation.

\subsection{Constraining \sseven  solutions\label{sub:Constraining-S7-solutions}}

The approach of \cite{Grignani:2008is} (and others) to strings on
$CP^{3}$ is to find solutions on the sphere $S^{7}\in\mathbb{C}^{4}$,
and then further demand that the two Noether charges from $\partial_{\gamma}$
vanish:\[
0=C_{0}\equiv\sum_{i=1}^{4}\im\left(\bar{Z}_{i}\partial_{t}Z_{i}\right)\:,\qquad0=C_{1}\equiv\sum_{i=1}^{4}\im\left(\bar{Z}_{i}\partial_{x}Z_{i}\right).\]
This is true for the $RP^{2}$ solution \eqref{eq:GHO-magnon-Z} given
by \cite{Grignani:2008is}, and more generally, for any solution on
the larger $RP^{3}$ subspace of section \ref{sub:Subspace-RP3}.
In terms of the co-ordinates $\vec{w}$ from \eqref{eq:rotate45-defn-w},
the condition $w_{3}=w_{4}=0$ which defines this subspace implies
$C_{0}=C_{1}=0$, and also reduces the equations of motion \eqref{eq:eq-of-m-in-Z}
to those for the sphere $S^{3}$ embedded in $(w_{1},w_{2})$. 

But more general solutions, such as the $CP^{1}$ solution \eqref{eq:GGY-magnon-Z},
do not solve these constraints, nor do they solve the equations of
motion for $S^{7}\subset\mathbb{C}^{4}$. So these conditions (solution
on $S^{7}$, and $C_{0}=C_{1}=0$) are certainly not necessary for
a solution. Whether they are sufficient is not entirely clear to us.%
\footnote{A similar approach to strings on the sphere is to find solutions in
flat (embedding) space and then reject all those which do not have
$\rho=1$. In this case solving the flat space equations and having
$\rho=1$ is sufficient to find a solution, but not necessary. 

For example, when studying loops of string rotating in $S^{3}$, there
is one critical speed at which they are solutions in unconstrained
$\mathbb{R}^{4}$ too. \cite{Roiban:2006jt} But faster and slower
motions are possible on the sphere, with extreme cases of a point
particle and a stationary hoop, which are not solutions in $\mathbb{R}^{4}$.%
} 

We noted in section \ref{sub:Subspace-RP3} that when working in the
subspace $RP^{3}$, the second term in the definition of charges $J_{i}$
\eqref{eq:Defn-Jz1-with-rho} vanishes, and what is left is the definition
of the conserved charge from rotational symmetry of the $z_{i}$ plane
one would expect in $S^{7}$. Here we can add that the term which
vanishes is $\left|Z_{i}\right|^{2}C_{0}/\rho^{4}$. This does not
vanish for the $CP^{1}$ case \eqref{eq:GGY-magnon-Z}, see footnote
\ref{fn:Note-on-wrong-J-defn}.

Finally, we note that in terms of charges $J_{i}$ we used throughout,
something like the constraint $C_{0}=0$ does hold: $\sum_{i=1}^{4}J_{i}=0$
follows trivially from the definition \eqref{eq:Defn-Jz1-with-rho}.

\begin{small}\bibliographystyle{my-JHEP-3}
\bibliography{/Users/me/Documents/Papers/complete-library-processed,complete-library-processed}

\end{small}
\end{document}